\documentclass{moriond}

\usepackage{amssymb,amsmath}
\usepackage{xspace}
\usepackage{graphicx}   
\usepackage{units}

\def\Journal#1#2#3#4{{#1} {\bf #2}, #3 (#4)}

\def\NIMA{{\em Nucl. Instrum. Methods} A}
\def\NPA{{\em Nucl. Phys.} A}
\def\NPB{{\em Nucl. Phys.} B}
\def\PLB{{\em Phys. Lett.}  B}
\def\PRL{\em Phys. Rev. Lett.}
\def\PRD{{\em Phys. Rev.} D}
\def\PPNP{{\em Prog. Part. Nucl. Phys.}}
\def\NPPS{{\em Nucl. Phys. Proc. Suppl.}}

\def\etal{{\em et al.} }

\def\be{\begin{equation}}
\def\ee{\end{equation}}
\def\bea{\begin{eqnarray}}
\def\eea{\end{eqnarray}}

\newcommand{\minerva}{MINERvA\xspace}
\newcommand{\minos}{MINOS\xspace}
\newcommand{\numu}{\ensuremath{\nu_{\mu}}\xspace}
\newcommand{\numubar}{\ensuremath{\bar{\nu}_{\mu}}\xspace}

\newcommand{\GeV}{\ensuremath{\mbox{GeV}}\xspace}

\newcommand{\GeVc}{\ensuremath{\mbox{GeV}}\xspace}

\newcommand{\Xbj}{\ensuremath{x}\xspace}
\newcommand{\Enu}{\ensuremath{E_{\nu}}\xspace}

\newcommand{\Emu}{\ensuremath{E_{\mu}}\xspace}
\newcommand{\thetamu}{\ensuremath{\theta_{\mu}}\xspace}

\newcommand{\recoilE}{\ensuremath{\nu}\xspace}
\newcommand{\Qsq}{\ensuremath{Q^{2}}\xspace}

\newcommand{\Photo}{\includegraphics[height=35mm]{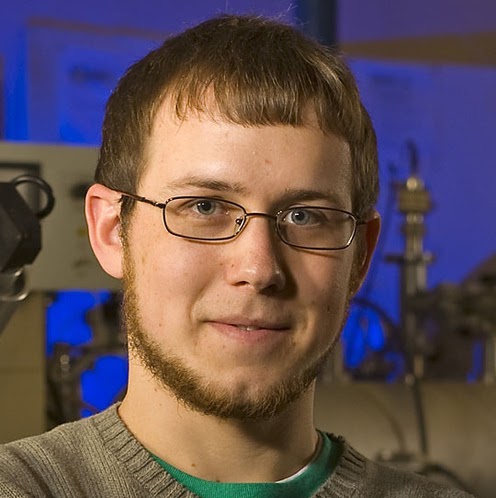}}

\begin{document}
\vspace*{4cm}
\title{Nuclear Target Cross Section Ratios at MINERvA}

\author{ Brian G. Tice\textit{, on behalf of the MINERvA Collaboration}}

\address{Argonne National Laboratory, Argonne, IL 60439, USA \footnote{Work performed at Rutgers, The State University of New Jersey, Piscataway, New Jersey 08854, USA.}}

\maketitle\abstracts{
  Measurements of \numu inclusive charged-current cross section ratios on carbon, iron, and lead relative to scintillator are presented.
  Data for the analysis were collected by the fine-grained \minerva detector in the NuMI beamline at Fermilab.
  This is the first direct measurement of nuclear dependence in neutrino scattering.
  The ratios show a depletion at low Bjorken $x$ and enhancement at large $x$, both of which increase with the nucleon number of the target.
  The data exhibit trends not found in GENIE, a standard neutrino-nucleus event generator, or alternative models of nuclear modification to inelastic structure functions.
}

\section{Introduction}
The European Muon Collaboration (EMC) made the landmark observation that the nucleon structure function $F_{2}$ is effectively altered when the nucleon is bound in a nucleus~\cite{Aubert.1983xm,Aubert1987740}.
EMC measured $F_{2}$ using both deuterium and iron using identical experimental conditions, and found that the ratio $F^{Fe}_{2}/F^{D}_{2}$ differs from unity as a nontrivial function of Bjorken's dimensionless scaling variable $x$.
This behavior has been extensively verified using charged lepton deep inelastic scattering but not directly measured with neutrino scattering~\cite{arneodo,geesaman1995nuclear,norton,rith}.
Despite vigorous experimental and theoretical work, the origins of this so-called ``EMC effect'' remain unknown. 
Measuring the effect with neutrinos is a crucial missing piece in understanding the origins of this effect.

Neutrino scattering differs from that of charged leptons in that it involves the axial-vector current.
For this reason, the nuclear dependence of $F_{2}$ observed in neutrino scattering is expected to differ from that of charged lepton scattering.
The neutrino scattering cross section has a contribution from an additional inelastic structure function $F_{3}$, which is not relevant in charged lepton scattering.
Although nuclear effects in neutrino and charged lepton scattering are expected to be different, the effects observed in charged lepton scattering are applied directly to neutrino interaction models.

\section{Experimental Procedure}

\subsection{MINERvA}
MINERvA (\textbf{M}ain \textbf{IN}jector \textbf{E}xpe\textbf{r}iment \textbf{$\nu$-A})~\cite{minerva.nim} is a neutrino scattering experiment at Fermi National Accelerator Laboratory (FNAL) in the NuMI beam.
NuMI delivers an intense flux of neutrinos or antineutrinos, created by the decay of mesons produced in $p$C collisions at \unit[120]{\GeVc}.
The \minerva detector employs fine-grained polystyrene scintillator (CH) for tracking and calorimetry.
The detector is built of 120 hexagonal modules stacked along the beam direction\footnote{The beam is directed at a downward angle of \unit[58]{mrad} with respect to the longitudinal detector axis.}.
The hexagonal main core of the detector is approximately \unit[5]{m} long and has inner and outer regions.
The inner detector (ID) is longitudinally organized into four subdetectors: the nuclear targets region, the fully active tracking region, downstream electromagnetic calorimetry (ECAL), and downstream hadronic calorimetry (HCAL).
The outer detector (OD) is a shell of hadronic calorimetry which surrounds and physically supports the ID.
The MINOS near detector sits downstream of \minerva and serves as a toroidal muon spectrometer.

\begin{figure}[h]
\centering
  \includegraphics[width=.45\textwidth]{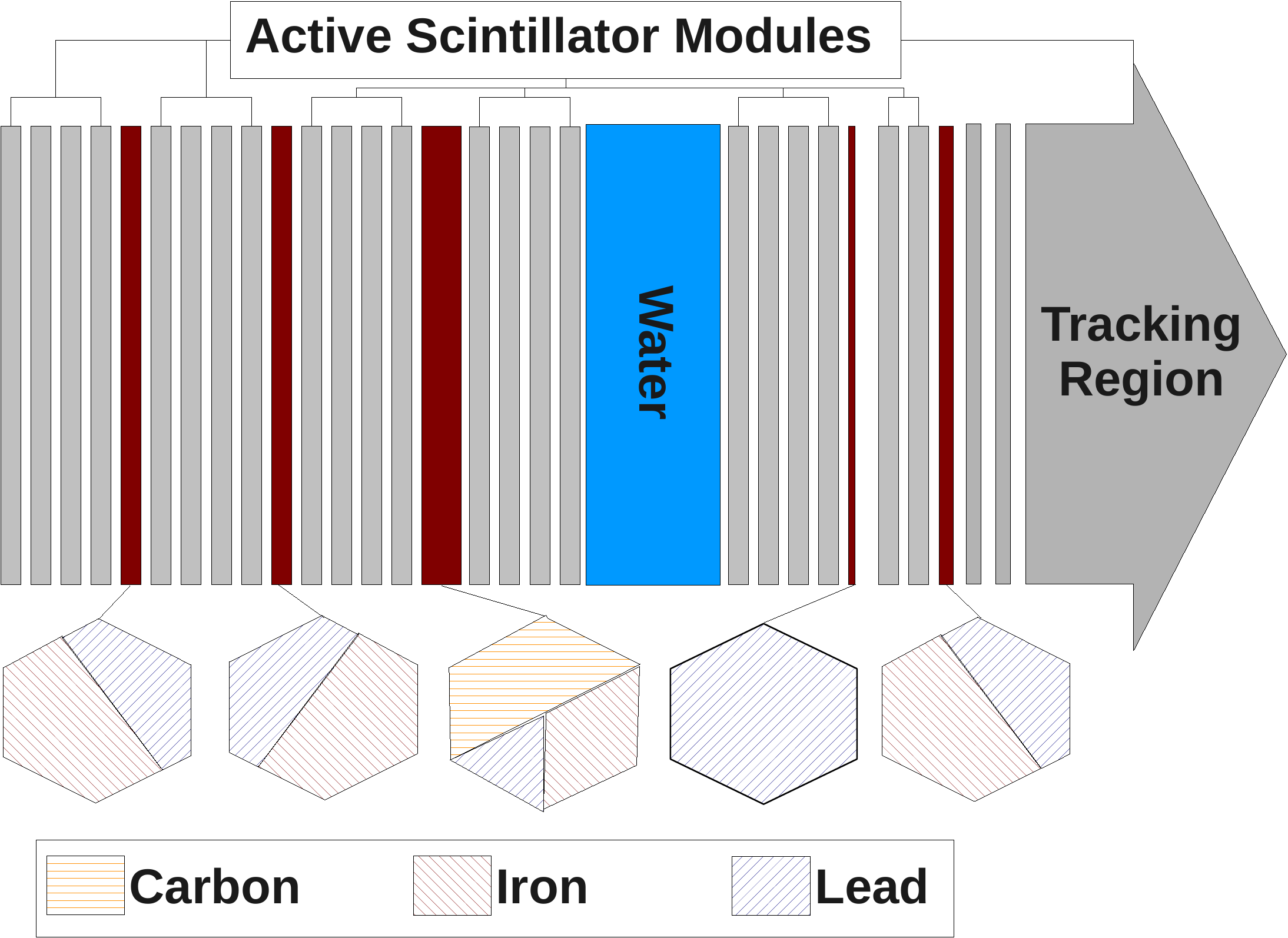}
  \caption{Schematic of the nuclear targets of \minerva.  Not shown is the liquid helium cryogenic vessel which sits just upstream (left) of the nuclear target region. The tracking region of the \minerva detector is immediately downstream (right) of the nuclear target region.}
  \label{fig:nucl_target_diagram}
\end{figure}

The nuclear targets region contains 22 tracking modules and 5 solid passive targets.
There are four tracking modules between targets, which is adequate for reconstructing tracks and showers.
A schematic of the nuclear targets region is shown in Fig.~\ref{fig:nucl_target_diagram}.
Passive targets are numbered upstream to downstream 1--5.
The targets are built by joining slabs of pure C, Fe, and Pb so that they occupy different transverse areas.

\subsection{Simulation of Nuclear Effects}
Neutrino events are simulated using the GENIE neutrino event generator~\cite{genie}.
Nuclei are modeled as a relativistic Fermi gas~\cite{smith1972neutrino} with high momentum tails in the nucleon momentum distribution~\cite{Bodek.1980ar,Bodek.1981wr}.
Pauli blocking is implemented by disallowing quasi-elastic events which produce a nucleon in the final state that does not have a momentum greater than the Fermi momentum.
A parameterization~\cite{Bodek.2004pc} to charged lepton measurements of the ratio $F_{2}^{Fe}/F_{2}^{(p+n)}$ is applied to all structure functions as a function of \Xbj, independent of \Qsq and A.

Two improved models of the modification of structure functions are also considered and compared to GENIE.
One model is an updated parameterization for ratios of $F_{2}^{A}/F_{2}^{(p+n)}$ which provides predictions specific for C, Fe, and Pb~\cite{by13}.
The other is the Kulagin-Petti microphysical model, which applies A-dependent corrections to neutrino-nucleon structure functions~\cite{PhysRevD.76.094023,Kulagin2006126}.
Despite predictions that vary up to 20\% for the absolute value of structure functions, these models agree in their predictions of structure function ratios to within 1\%.

\subsection{Analysis}
This analysis of charged-current \numu events uses data collected from $2.94\times 10^{20}$ protons on target taken between March 2010 and April 2012 when the beamline produced a broadband neutrino beam peaked at \unit[3.5]{\GeV} with $>95\%$ \numu at the peak energy~\cite{this}.
Charged-current events from passive targets 2--5 and the tracking region are considered.
The fiducial masses of C, Fe, Pb, and CH are 0.16, 0.63, 0.71, and 5.48 metric tons, respectively.
Events are required to have a reconstructed muon, which is identified as a track that exits \minerva and is matched to a track in the \minos near detector.
The kinematics are limited to $2\: \GeV < \Enu < 20\: \GeV$ and $\thetamu < 17^{\circ}$ in order to reduce the need for model-dependent acceptance corrections and eliminate backgrounds from \numubar interactions.

The energy \Emu and angle \thetamu of the muon come from fitted track parameters.
The energy of the hadronic system \recoilE is reconstructed as the calorimetric sum of all hits not associated with the muon that are recorded between \unit[20]{ns} before and \unit[35]{ns} after the interaction time.
Kinematic variables are calculated from these reconstructed quantities: 
$\Enu=\Emu+\recoilE$, 
$\Qsq=4\Enu\Emu\sin^{2}(\frac{\thetamu}{2})$, 
$\Xbj=\frac{\Qsq}{2M_{N}\recoilE}$,
where \Qsq is the negative of four-momentum transfer squared and $M_{N}$ is the average of proton and neutron masses.
Distributions of \Enu are corrected for finite resolution and detector smearing through iterative Bayesian unfolding~\cite{DAgostini.1994zf} informed by the generated \Enu values from GENIE.
Distributions of \Xbj are not unfolded to avoid introducing dependence on the nuclear model and because the smearing in \Xbj is significant.

The nucleus with which the neutrino interacted is inferred from the vertex position.
The vertex position is reconstructed using a Kalman filter when possible ($\sim$15\% of the time) and is otherwise taken to be the most upstream energy deposition.
Fitting the vertex is not possible for events with only one reconstructed track; for such events the vertex appears to be in the scintillator module adjacent to the passive target.
An event is identified with C, Fe, or Pb if its vertex is in the passive target or in an adjacent module.
The vertex must be \unit[2.5]{cm} away transversely from seams that join different materials in a single target.
Events in CH must have an interaction vertex in the tracker fiducial volume.
After all selection criteria, 5953 events in C, 19024 in Fe, 23967 in Pb, and 189168 in CH are selected for the analysis.

The passive target event selection admits a background of ``CH contamination'' from the surrounding CH modules of 20--40\%, which is subtracted through a data-driven procedure.
Kinematic spectra of events recorded in the large CH tracking volume are used to predict and subtract this background.
One weight $w^{t,A}(\Enu,\thetamu)$ is derived from a single particle muon simulation to account for the difference in geometry acceptance between the tracking volume and the nuclear targets region, where $t$ denotes target number 2--5 and $A$ stands for C, Fe, or Pb.
Another weight $w^{t,A}(\recoilE)$ is derived from GENIE simulation to account for differences in tracking efficiency due to differences in the way the hadronic system develops on the two regions of the detector.
Small backgrounds from neutral current ($<0.1\%$), \numubar ($<0.4\%$), and wrongly-assigned interaction nucleus events ($\sim0.5\%$) are estimated using simulation and subtracted.
\begin{figure}[ht]
\centering
\includegraphics[width=0.45\columnwidth]{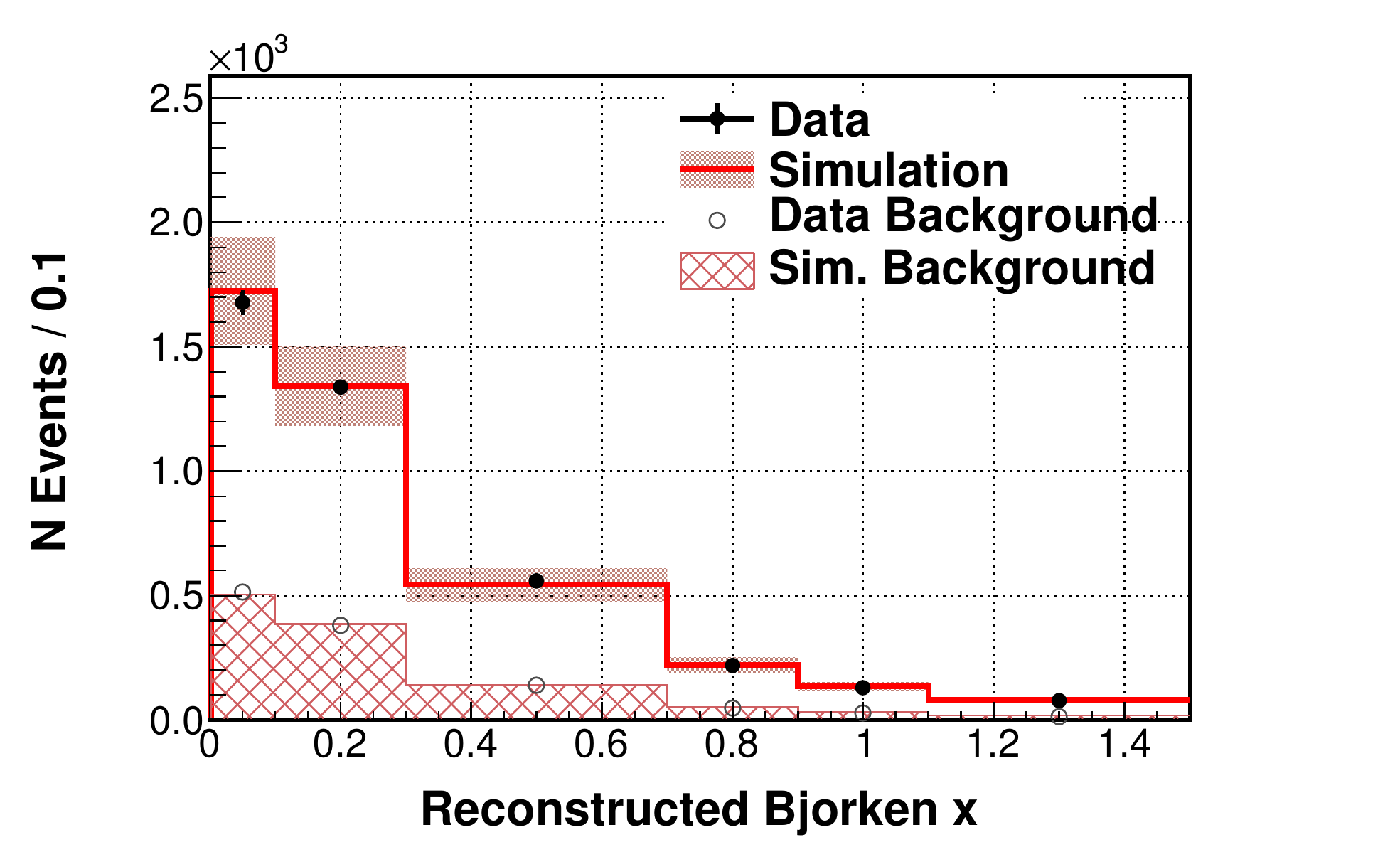}
\includegraphics[width=0.45\columnwidth]{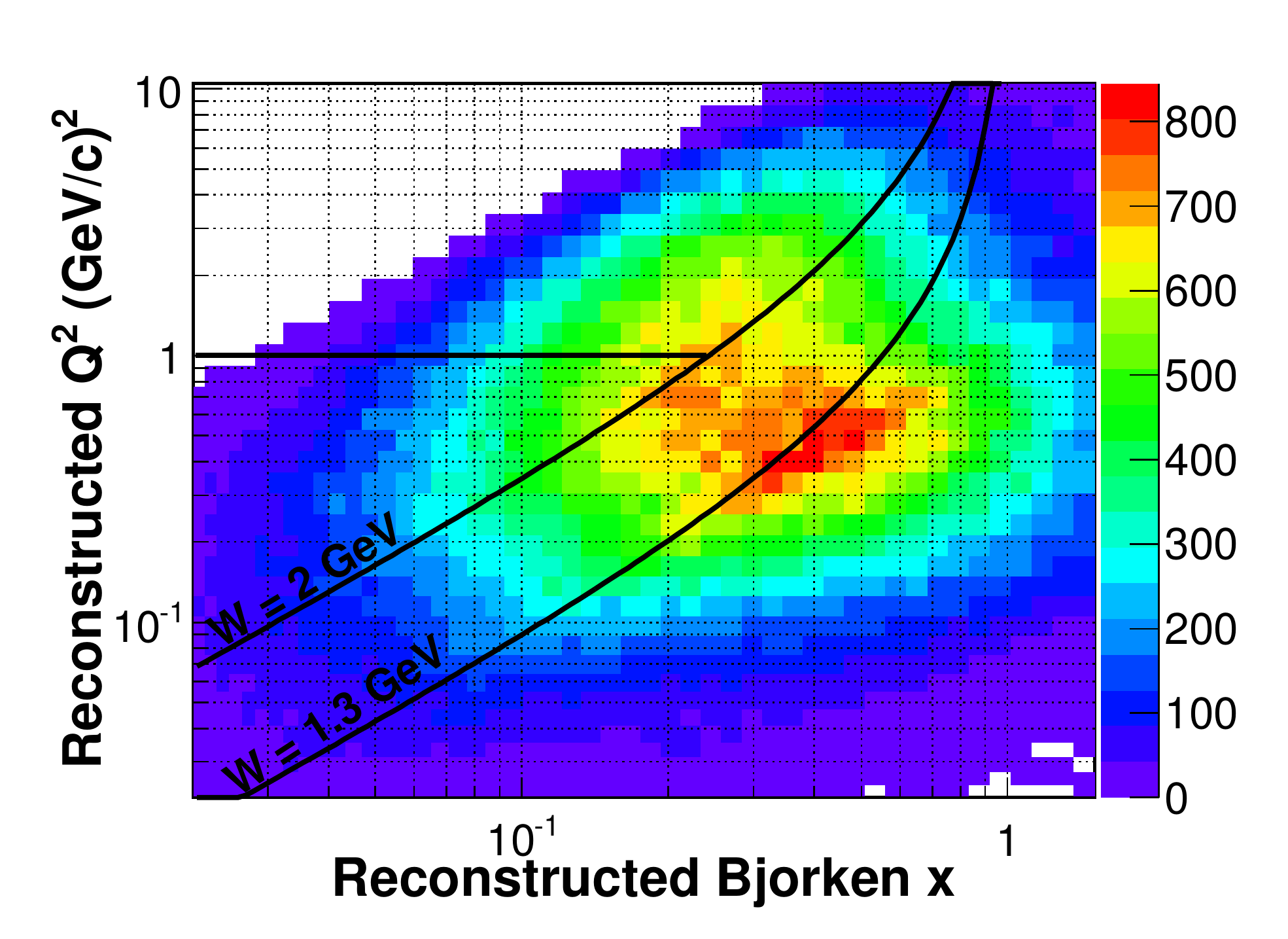}
\caption{ (Left) Reconstructed Bjorken $x$ distributions in data and simulation for selected inclusive \numu events in the iron of Target 2. The plot includes CH contamination separately estimated using data and simulated events in the tracker region.  Both simulation distributions are normalized to the data by the number of events passing all event selection criteria.  Events are scaled to a bin size of 0.1.  Events with \Xbj greater than 1.5 are not shown. (Right) Reconstructed \Qsq and \Xbj for all events used in the analysis with lines of constant invariant mass $W$ and \Qsq.}
\label{fig:sample}
\end{figure}
Figure~\ref{fig:sample} shows the reconstructed \Xbj distribution for events selected in the Fe of Target 2 and the estimated background from CH contamination.
Also shown is the kinematic space in \Xbj-\Qsq populated by all events used in the analysis.

\section{Results and Conclusions}
\begin{figure}[ht]
\centering
\includegraphics[trim = 1mm 00mm 18mm 8mm, clip, width=0.3\columnwidth]{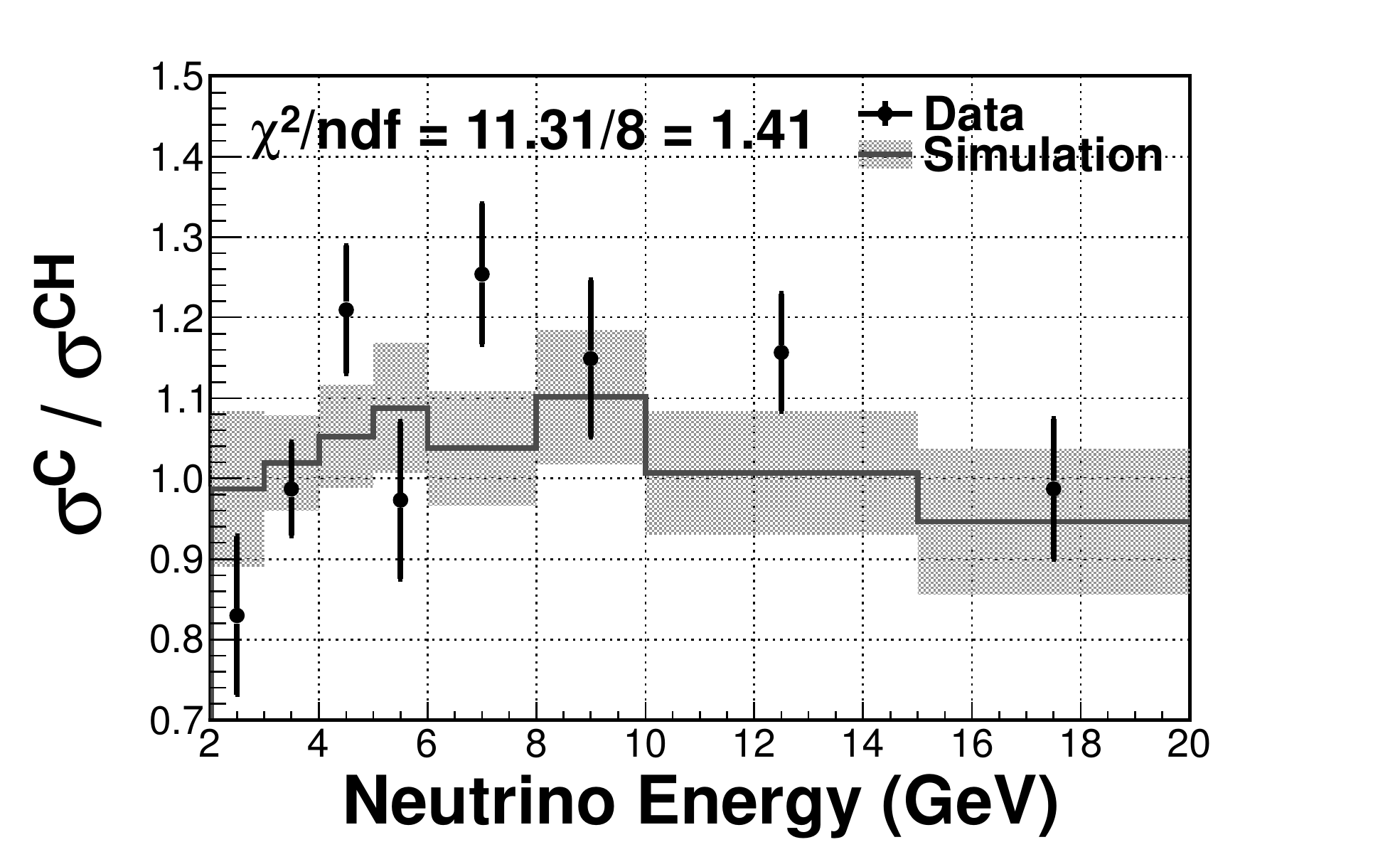}
\includegraphics[trim = 1mm 00mm 18mm 8mm, clip, width=0.3\columnwidth]{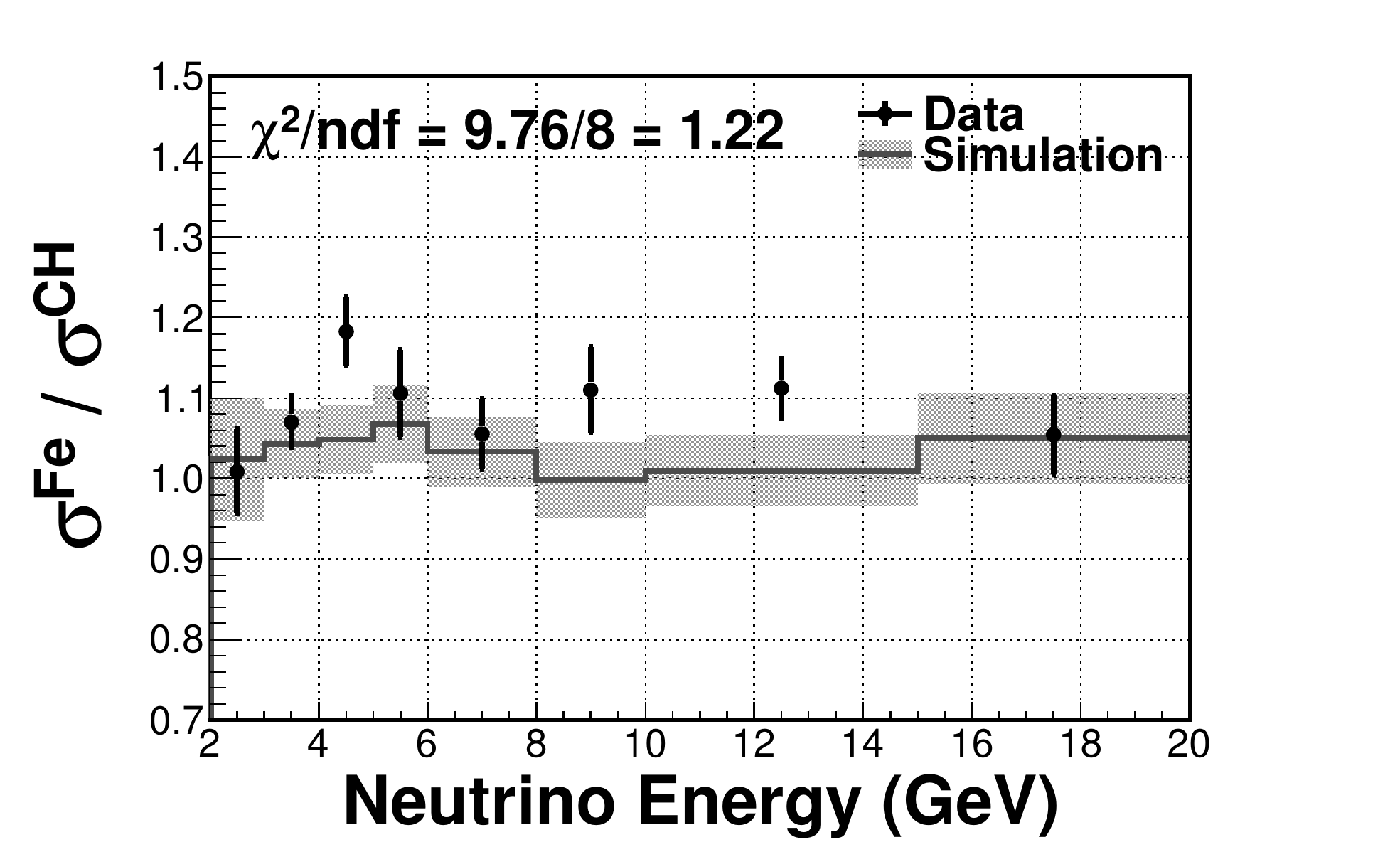}
\includegraphics[trim = 1mm 00mm 18mm 8mm, clip, width=0.3\columnwidth]{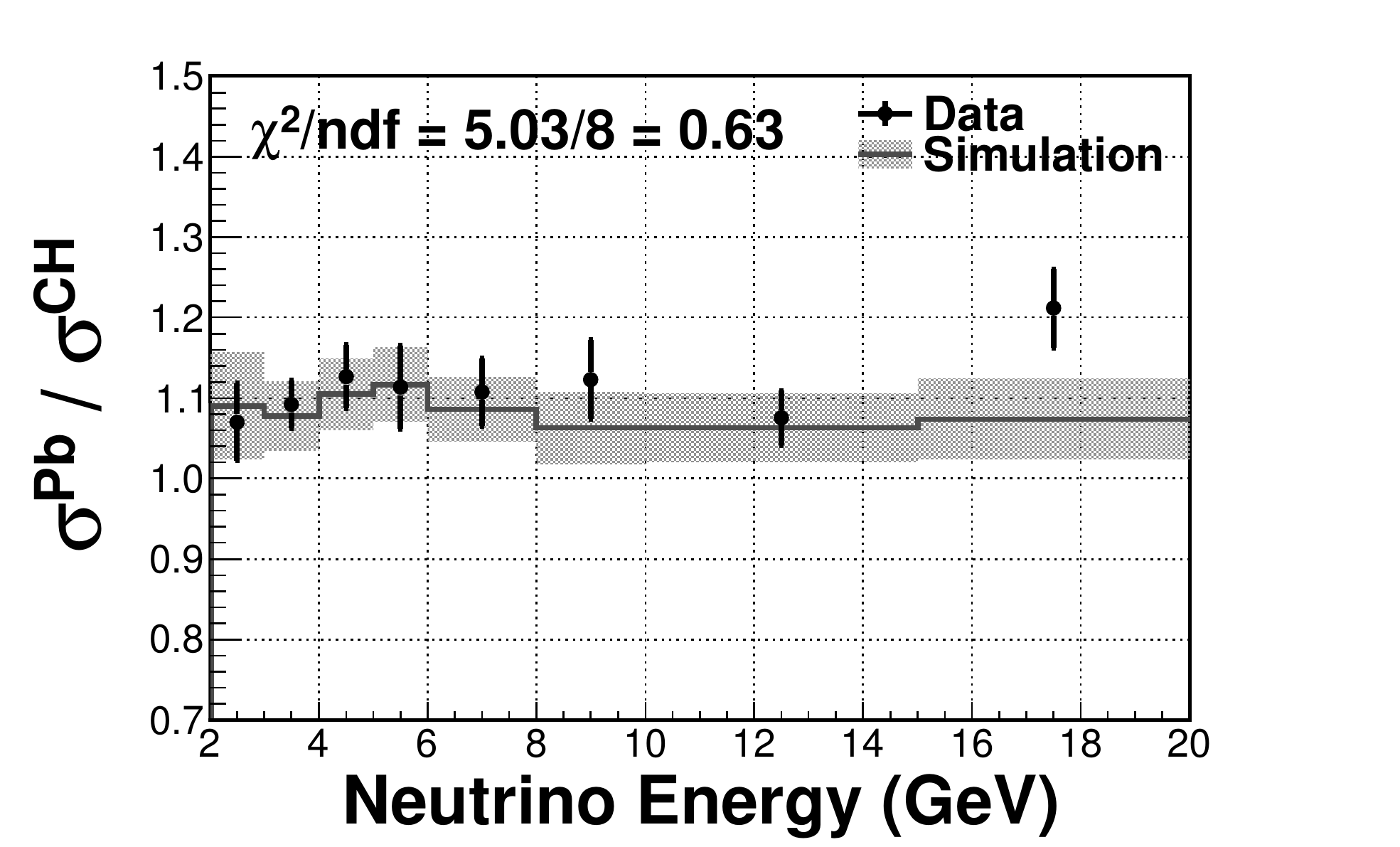}
\includegraphics[trim = 1mm 00mm 18mm 8mm, clip, width=0.3\columnwidth]{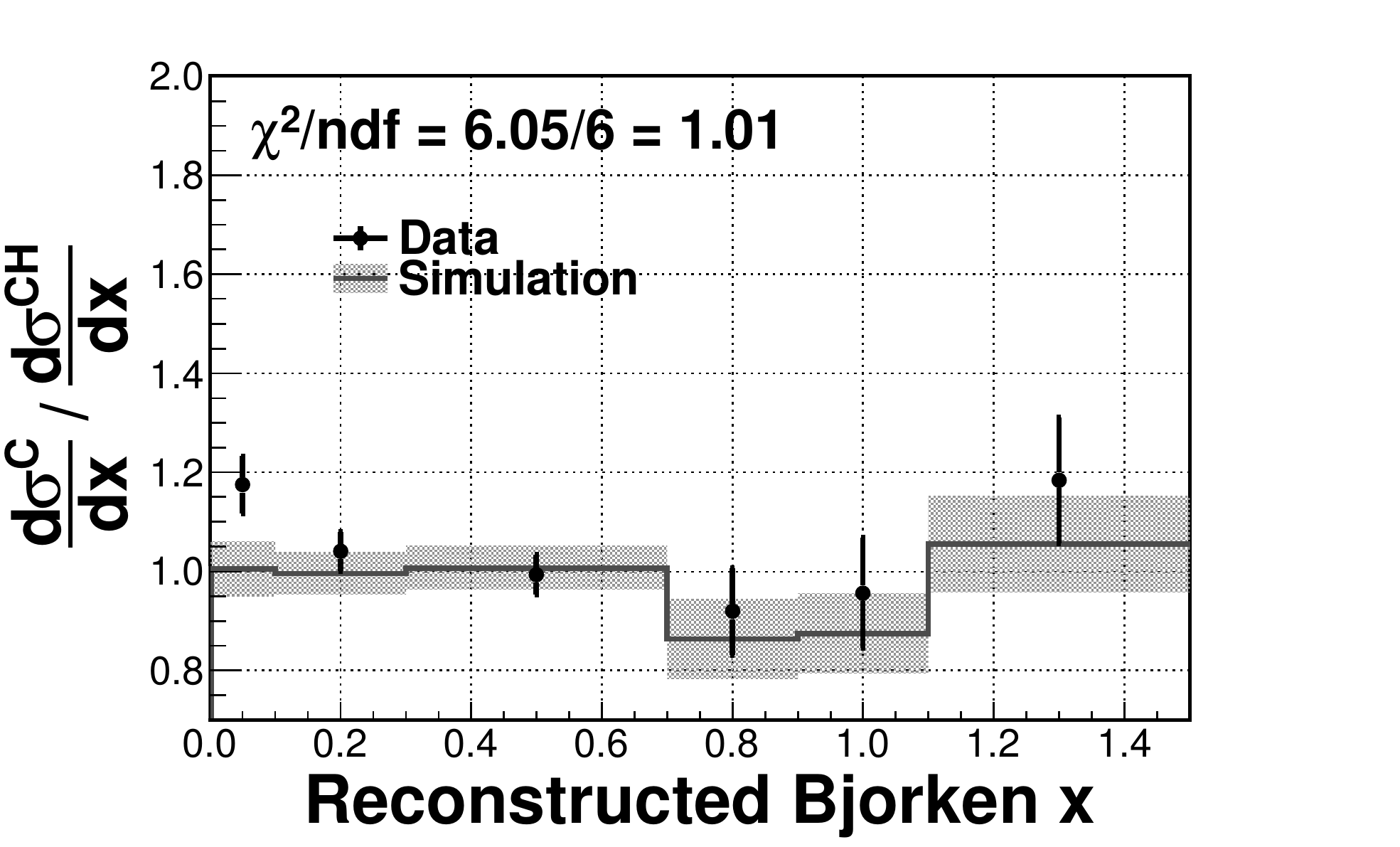}
\includegraphics[trim = 1mm 00mm 18mm 8mm, clip, width=0.3\columnwidth]{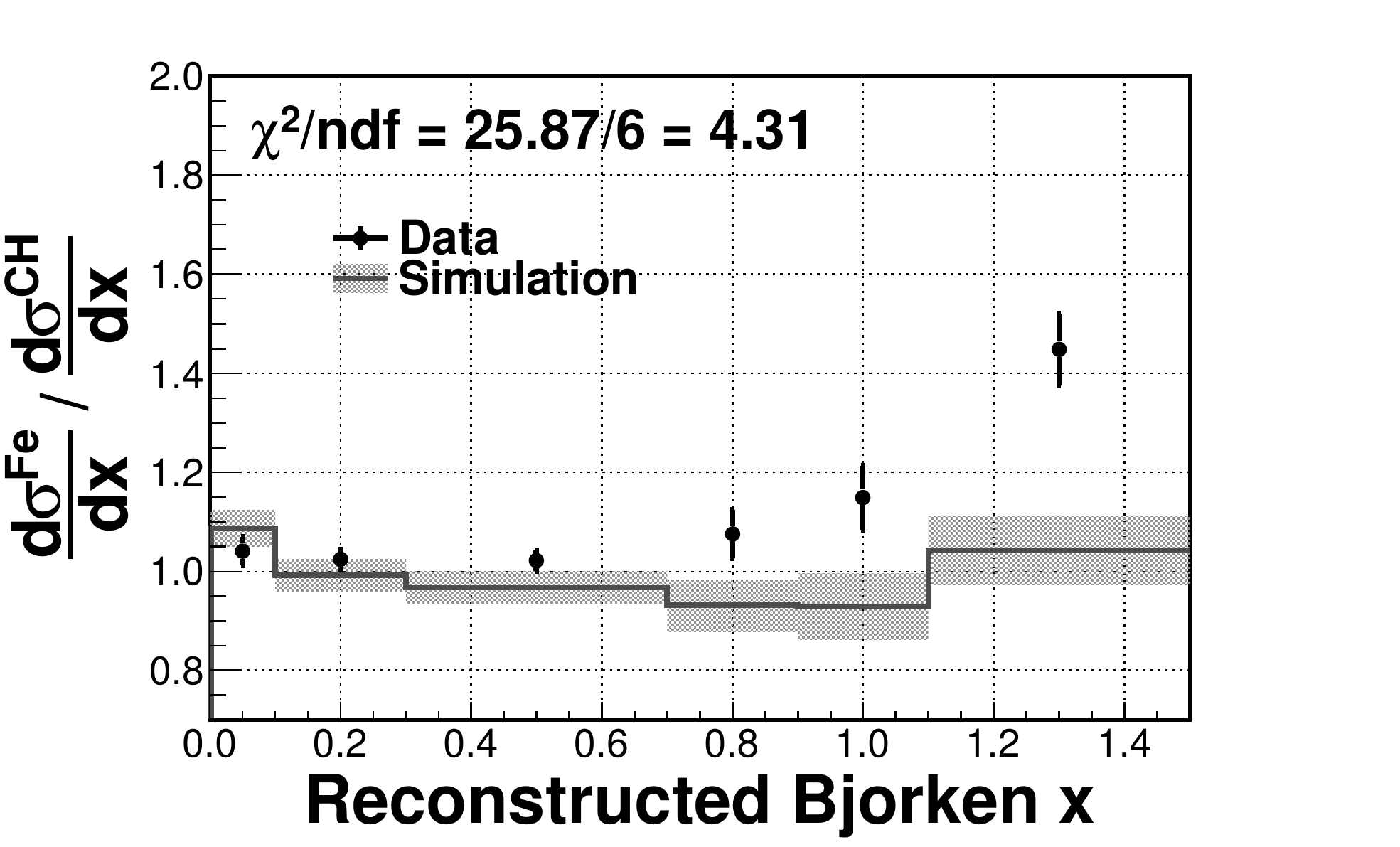}
\includegraphics[trim = 1mm 00mm 18mm 8mm, clip, width=0.3\columnwidth]{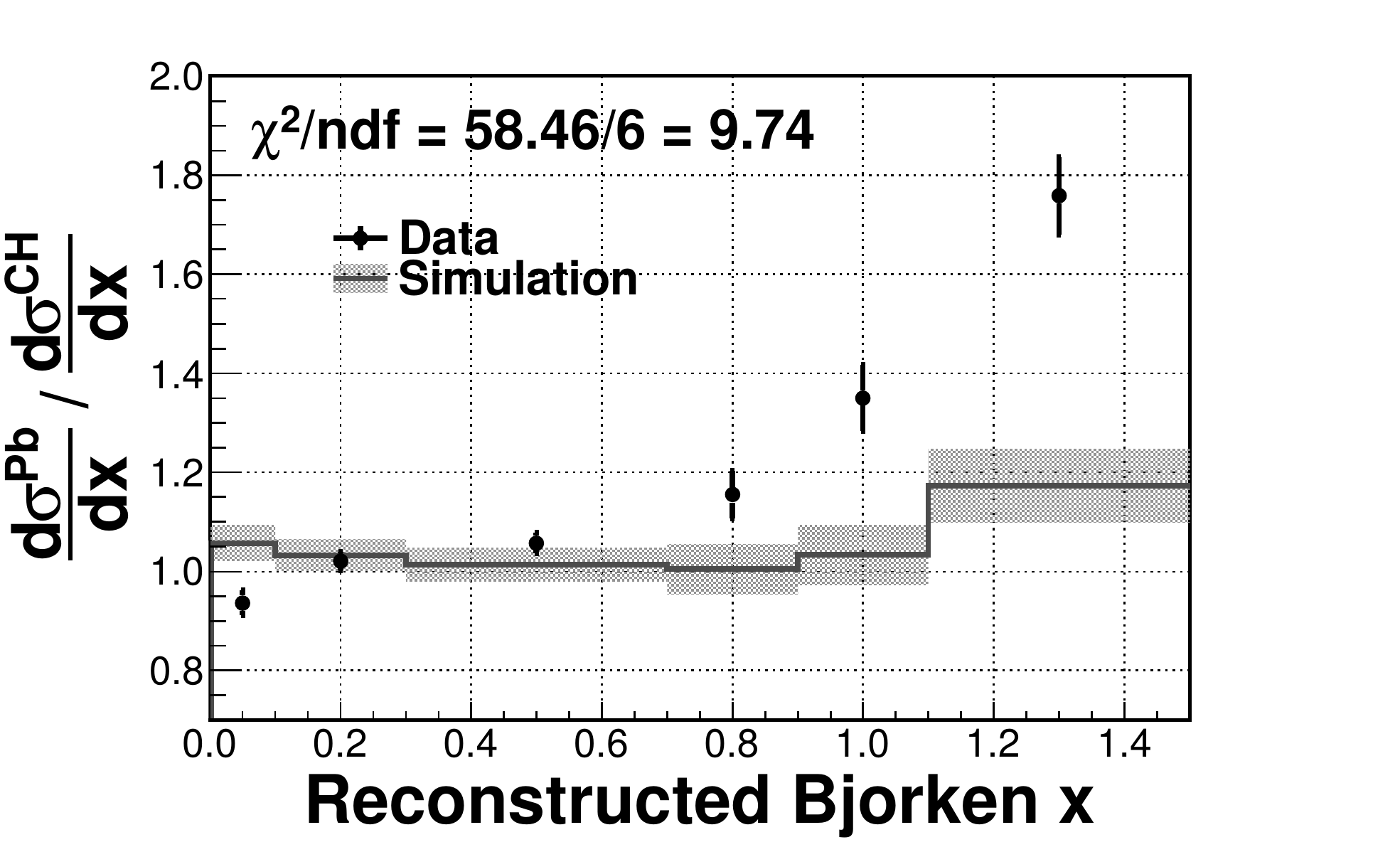}
\caption{Ratios of the charged-current inclusive \numu cross section per nucleon as a function of \Enu (top) and as a function of reconstructed \Xbj (bottom)
for C/CH (left), Fe/CH (middle), and Pb/CH (right).  Error bars on the data (simulation) show the statistical (systematic) uncertainties.  The $\chi^{2}$ calculation includes correlations among all bins shown.  Events with \Xbj greater than 1.5 are not shown.}
\label{fig:xsec_ratio_plots}
\end{figure}

Figure~\ref{fig:xsec_ratio_plots} shows measured ratios of the inclusive \numu cross section per nucleon with comparisons to GENIE simulation.
No isoscalar correction is applied.
For cross section ratios as a function of \Enu ($\sigma^{A}/\sigma^{CH}$), the simulation reproduces the data to within the limits of the measurement.
However, ratios as a function of \Xbj ($\frac{d\sigma^{A}}{d\Xbj}/\frac{d\sigma^{CH}}{d\Xbj}$) exhibit a depletion at low \Xbj and enhancement at large \Xbj growing with the nucleon number of the target nucleus, neither of which is reproduced by available simulations.

At lower \Xbj values, shadowing and antishadowing effects modify the structure function $F_{2}$~\cite{kopeliovich2012nuclear}.
Shadowing is understood to be caused by the phenomenon of quark multiple scattering, which produces a suppression at $\Xbj \lesssim 0.07$.
The mechanism causing the enhancement of $F_{2}$ in the antishadowing region $0.07 \lesssim \Xbj \lesssim 0.3$ is not known.
These effects are expected to be larger at low \Qsq.
The data presented here are predominately in the non-perturbative range at low \Qsq (80\% of events have $\Qsq<1~\GeV^2$), whereas the simulation is tuned to higher \Qsq charged lepton data.

At higher \Xbj values, the sample is dominated by quasi-elastic events ($\textgreater63\%$).
Since most neutrino oscillation experiments utilize quasi-elastic events, the inability to model nuclear dependence in neutrino scattering in this region is troubling.
T2K~\cite{PhysRevLett.111.211803,PhysRevLett.112.061802} uses C for a near detector and H$_{2}$O for the far detector, and so must use a model of nuclear dependence to do an extrapolation of event rates.
LBNE~\cite{lbne_new} must infer the nuclear effects in Ar from existing data on C, Fe, Pb.
The results presented here indicate that the models of nuclear dependence in neutrino scattering are not adequate for the precision of modern neutrino experiments.

\end{document}